**Strengthening science, technology, and innovation-based incubators to help achieve Sustainable Development Goals: Lessons from India**

Kavita Surana[1*], Anuraag Singh[2], Ambuj Sagar[2,3]

[1]School of Public Policy, University of Maryland, College Park, USA

[2] DST Centre for Policy Research, Indian Institute of Technology Delhi, India

[3] School of Public Policy, Indian Institute of Technology Delhi, India

*corresponding author

## Abstract

Policymakers in developing countries increasingly see science, technology, and innovation (STI) as an avenue for meeting sustainable development goals (SDGs), with STI-based startups as a key part of these efforts. Market failures call for government interventions in supporting STI for SDGs and publicly-funded incubators can potentially fulfil this role. Using the specific case of India, we examine how publicly-funded incubators could contribute to strengthening STI-based entrepreneurship. India's STI policy and its links to societal goals span multiple decades—but since 2015 these goals became formally organized around the SDGs. We examine why STI-based incubators were created under different policy priorities before 2015, the role of public agencies in implementing these policies, and how some incubators were particularly effective in addressing the societal challenges that can now be mapped to SDGs. We find that effective incubation for supporting STI-based entrepreneurship to meet societal goals extended beyond traditional incubation activities. For STI-based incubators to be effective, policymakers must strengthen the 'incubation system'. This involves incorporating targeted SDGs in specific incubator goals, promoting coordination between existing incubator programs, developing a performance monitoring system, and finally, extending extensive capacity building at multiple levels including for incubator managers and for broader STI in the country.

**Keywords:** Science technology and innovation; developing countries; entrepreneurship; incubators; public policy; sustainable development goals

## 1. Introduction

Policymakers and scholars—especially in developing countries—have long perceived that innovation and entrepreneurial activity can potentially generate new ways to advance economic development, employment, and more effective and efficient delivery of services (Acs et al., 2008; Audretsch et al., 2007; Naudé, 2010; OECD, 2013; Szirmai et al., 2011). A key focus of public policies for strengthening science, technology, and innovation (STI) in developing countries has been through promoting entrepreneurship and supporting such startups through public incubators—traditionally defined as formally organized entities formed with governmental support that offer various services for the conversion of individual ideas from high-risk innovation and early-stage startup incubatees to more advanced, market-oriented enterprises. Countries like India, Brazil, and China have had long-running public incubator programs aimed at advancing economic and social development while countries elsewhere, e.g. in Africa, have recently started to expand STI-based incubator activity (Akçomak, 2009; Chandra and Fealey, 2009; Chandra and Silva, 2012; Dalmarco et al., 2018; Lalkaka, 2002; Scaramuzzi, 2002; Tang et al., 2013; The Economist, 2017).

While developing countries have had a range of economic, social, and environmental objectives for several decades, the 2030 Agenda and the adoption of the Sustainable Development Goals (SDGs) in 2015 formalized these goals globally and in that process, cemented the importance of STI and entrepreneurship in the SDGs and their implementation (UNFCCC, 2018)[1]. The broader recognition of the role of STI for SDGs is translating to an exploration of concrete actions on startup incubators and entrepreneurial activity that incubators can help promote to implement these goals.[2] Given the renewed emphasis of incubators and startups in the context of SDGs, questions emerge for policymakers on how publicly-funded incubators in developing countries can most effectively link STI-based entrepreneurship with multiple SDGs.

[1] *Goal 8: Decent work and economic growth* target 8.3 mentions promoting "development-oriented policies that support productive activities, decent job creation, entrepreneurship, creativity and innovation…" (United Nations, 2015). *Goal 9:Industries, innovation, and infrastructure* specifically aims to enhance innovation and research and development activities in target 9.5 (United Nations, 2015). Similarly, *Goal 13: Climate action* links to the Paris Agreement; the latter states that "Accelerating, encouraging and enabling innovation is critical for an effective, long-term global response to climate change and promoting economic growth and sustainable development" (UNFCCC, 2015; United Nations, 2015).

Despite the significance of incubators in linking STI with SDGs, there is a gap in the understanding of incubators' goals, activities, and their contributions to developmental outcomes in developing countries. Extant literature has extensively analyzed incubators and their impact on innovation and entrepreneurship largely in European countries, the United States, or other industrialized countries (see, for example, Dutt et al., 2015; Kochenkova et al., 2016; Mian et al., 2016; Phan et al., 2005), noting that there is no single or consistent framework of assessment to measure 'success' (Phan et al., 2005). Studies on developing countries are few—most studies analyze incubator activity under broader assessments of STI and entrepreneurship (Acs et al., 2008; Akçomak, 2009; Autio et al., 2014; Naude, 2013; Naudé, 2010) while only a few studies analyze what incubators have actually accomplished (Akçomak, 2009; Lalkaka, 2002; Tang et al., 2013). There is a substantial distinction between the traditional incubators in industrialized country contexts and the incubators in developing country contexts, though. The traditional incubators in industrialized countries act as intermediaries who complement other actors in what are often already strong innovation systems

[2] For example, the Green Climate Fund (GCF), the financial arm of the UNFCCC plans to support climate technology incubators and accelerators in developing countries. As part of this, the GCF, in partnership with the Technology Executive Committee (TEC) and the Climate Technology Centre and Network (CTCN) of the UNFCCC held a global dialogue in 2018 to advance climate technology incubators in developing countries (Green Climate Fund, 2018; UNFCCC, 2018)

and strong institutions to support and expand private-sector-driven technology innovation. While industrialized countries do have publicly-funded incubators to meet specific policy goals by supporting high-risk startups (e.g., Leyden and Link, 2015), they are still operating in a well-resourced and dynamic innovation context. In contrast, in developing countries, incubators have been a central element of STI policy extending their role beyond intermediary organizations. For many decades, incubators have focused on achieving the social, environmental, and economic goals now embodied in the SDGs. These SDG related goals are unlikely to be addressed by innovation systems that generally are characterized by resource constraints, weak innovation institutions, and sporadic dynamism at best in private and public actors. Multiple market and policy shortcomings—e.g., related to underdeveloped institutions, lack of human and financial resources, insufficient paying capacity by beneficiaries (Khanna and Palepu, 1997)—exacerbate the ability to develop new knowledge or to effectively translate it into application, leading to underinvestment by the private sector in innovation or in high-risk startups. At the same time, insufficient monetization of public goods (often central to the SDGs) further impedes private sector investment and activities in these areas. While literature on incubators has recognized the need for context-specific analysis of incubator goals and outcomes, literature on incubators in developing countries is disproportionately low despite the clear contrasts in the role of incubators in industrialized and developing countries.

This paper addresses the following research questions: to what extent have STI-based incubators and the startups that they support historically helped to implement the development objectives of the kind now organized around SDGs in developing countries? And how might they better help achieve the SDGs? To address these questions, we examine India's incubator experiences from three perspectives. First, focusing on the policy drivers and development goals for STI-based entrepreneurship, we assess *why* publicly-funded incubators were created. Second, focusing on the public agencies responsible for designing and implementing government-led programs for supporting STI-based entrepreneurship, we analyze *what* these agencies did to implement policies using incubators as intermediaries and what were the outcomes. Third, focusing on actual incubators, we analyze *how* incubators have been effective in the context of implementing societal goals that map to the SDGs. An analysis of these experiences then allows us to reflect on how these incubators could better help India achieve its SDGs, with the expectation that many of these lessons may also be useful for other developing countries.

While our analysis on the potential relevance of incubators for implementing SDGs focuses on India, we suggest that it is suitable for a case study because its incubator activities echo global trends—from its historical multi-decade experience in governmental STI-based entrepreneurship programs that led to the creation of over 140[3] incubators between 1985-2014, to renewed policy emphasis since the mid 2010s that called for supporting startups by massively expanding the number of existing incubators (alongside improving other factors such as easing regulatory barriers, providing high-risk funding for startups, and increasing innovation capacity)[4]. But despite three decades of continued experiences linked to incubators and whether they met societal, environmental, or economic development goals, limited systematic analyses exist of one of the longest running governmental efforts to promote STI-based entrepreneurship in a major developing country (GIZ, 2012; Lalkaka, 2002; Tang et al., 2014). As recent STI policies start to prioritize the expansion of incubators while being explicitly linked to the SDGs(UN India, 2019a, 2019b, p. 9), an analysis of India's rich past experiences can provide useful lessons for itself as well as other developing countries that have new and ambitious plans to mobilize STI-based entrepreneurship for implementing SDGs.

[3] Author calculations

[4] In 2014-2016, STI-based entrepreneurship became a specific policy priority and the government announced a set of ambitious national-level policies (e.g., Startup India, Atal Innovation Mission) and various regional-level initiatives for startups

The rest of the paper is structured as follows. Section 2 provides a brief background on public policy for STI-based entrepreneurship and incubators. Section 3 focuses on the case context and our approach for the analysis of incubators in India. Section 4 discusses our results on the policy motivations behind why incubators were created, what public agencies did to implement policy and what were the outcomes, and finally how incubators were able to implement societal goals that map to the SDGs. Section 5 highlights policy implications for incubators in India and other developing countries. Section 6 concludes.

## 2. Background

In this section, we highlight the literature on STI-based entrepreneurship and startups in developing countries (2.1), how incubators broadly support STI-based entrepreneurship (2.2), and finally, the specific role incubators play in developing countries (2.3).

### 2.1. STI policy and startups in developing countries

The importance of innovation and entrepreneurship in pushing economic development has been long established, with clear recognition of science and technology as precursors to innovation (Freeman and Soete, 1997; Naudé, 2010; OECD, 2013; Schumpeter, 1934; Szirmai et al., 2011). Consequently, governments have focused on generating and supporting STI-related activities expecting economic welfare outcomes such as employment and industrial competitiveness. STI policy (in industrialized as well as developing countries) has emphasized on three areas. One, strengthening the 'supply-side' for STI—e.g., by promoting science and technology-based education, setting up research and development (R&D) laboratories, funding R&D in universities, creating science and technology-based large public enterprises, improving intellectual property rights (IPR). (e.g., Etzkowitz and Leydesdorff, 2014; Fagerberg et al., 2005; Nelson, 1993). Two, supporting entrepreneurship at large—e.g., by implementing policies and programs that finance small and medium business or startups, easing regulatory barriers to start or end a business. (e.g., Acs and Szerb, 2007; Minniti, 2008). Three (discussed in more detail in 2.2), strengthening the links between STI, entrepreneurs, startups, and markets—e.g., through setting up incubators (and other intermediaries like science parks, technology transfer centers, etc.) that support technology transfer especially for technologies that would be unable to advance to market in the absence of different types of public support (e.g., Mian et al., 2016; Phan et al., 2005).

In the context of implementing SDGs, the link between STI and specific SDGs has become most apparent in the emergence of startups linked to the SDGs and the incubators that would support these (See Table 1). While startups have become attractive in industrialized and developing countries alike because of their perceived ability to be nimble and to quickly adapt to market needs (and therefore deliver quick results), STI-based startups for SDGs in developing countries face a unique set of interrelated challenges. *First*, while startups are risky by definition—the risks are amplified for STI product based startups (compared to service based startups) because of risks of technology failure, the unavailability of infrastructures to support technologies, the lack of technology-trained human capacity needed for success (Autio et al., 2014; Ghani et al., 2014). *Second*, the success of STI-based startups depends in multiple ways on the private sector. But because of multiple market failures in developing countries, the private sector has undervalued the long-term societal benefits of implementing SDGs and has underinvested in STI-based startups. These market failures include: (a) lack of resources to test or validate risky STI-based ideas, (b) paucity of early adopters willing to take up new technologies, (c) lack of long-term investment in STI-based startups (excluding IT-based startups) that clashes with the long timescales needed for STI-based startups to demonstrate outcomes (STI based startups may need more time to create or test prototypes, to manage supply chains and physical distribution of the product, and to demonstrate market acceptance), and (d) low monetary returns for public goods related to SDGs despite their high societal benefits (e.g., Khanna and Palepu, 1997). Overall, effective government-led activities that link STI-based startups with the implementation of SDGs would need to incorporate the unique context and challenges that such startups may face in developing countries.

**TABLE 1: Examples of the links between STI and SDGs from India and other developing countries, including some examples of incubators whose operations directly map to the SDGs.**

| Sustainable Development Goals | | Description | Illustrative examples of the growing links between STI and SDGs from India and other developing countries |
|---|---|---|---|
| **1** | No poverty | End poverty in all its forms everywhere | Mobile banking supports low income households get access to banking with low transactional costs. Startups like Kenya's M-Pesa operate in this space and have contributed to lifting households out of poverty (Suri and Jack, 2016). |
| **2** | Zero hunger | End hunger, achieve food security and improved nutrition and promote sustainable agriculture | Indigram Labs Foundation, an incubator in India, focuses on agriculture technology. Its incubatee startups include New Leaf Dynamic Technologies that is building off-grid cold storage units for farmers to help minimize post-harvest losses (Indigram Labs, 2018). |
| **3** | Good health and well-being | Ensure healthy lives and promote well-being for all at all ages | MicroMek is a startup based in Malawi that aims to develop low-cost autonomous drones to deliver medicines and health care supplies to remote populations (Kaliati, 2019) |
| **4** | Quality education | Quality education | Injini, a pan-African incubator, focuses on education technology. Its incubatee startups included M-Shule, a startup from Kenya, that aimed to develop an adaptive learning platform that could deliver personalized lessons through SMS (Injini, 2019). |
| **5** | Gender equality | Achieve gender equality and empower all women and girls | Solar Sister is a startup based in Sub-Saharan Africa and empowers women to become entrepreneurs by bringing clean energy into their communities through women-led enterprises (Solar Sister, 2019) |
| **6** | Clean water and sanitation | Ensure access to water and sanitation for all | India's Department of Industrial Policy and Promotion (DIPP) organized a "Grand Challenge" competition for startups that link to the government-led mission of clean water and sanitation (Swachh Bharat Mission). One of the winning startups, Altersoft Innovations, is developing smart, self-cleaning public toilets using internet of things (IoT) technologies (DIPP, 2018). |
| **7** | Affordable and clean energy | Ensure access to affordable, reliable, sustainable and modern energy | M-Kopa is a Kenya-based startup that sells solar home systems and allows buyers to make digital payments in a pay-as-you-go model (M-Kopa, 2019) |
| **8** | Decent work and economic growth | Promote inclusive and sustainable economic growth, employment and | In Uttar Pradesh (the most populous state in India), the state Information Technology (IT) and Startup policy aims to use IT as a way of bringing in economic growth and has established four incubators as part of this effort (Government of Uttar Pradesh, 2017). |
| | | decent work for all[5] | |
| **9** | Industry innovation and infrastructure | Build resilient infrastructure, promote sustainable industrialization and foster innovation[6] | Supporter by India's Biotechnology Industry Research Assistance Council (BIRAC), the Centre for Cellular and Molecular Platforms (C-CAMP) runs an incubator that focuses on supporting STI-based startups in the life sciences industry (C-CAMP, 2019) |
| **10** | Reduced inequalities | Reduce inequality within and among countries | The Assistive Technology Accelerator (ATA) based in India supports startups and persons with disability. Its incubatee startups include Eye-D that is developing an app to help visually impaired with travel, identification of objects, and reading text (Assistive Technology Accelerator, 2019) |
| **11** | Sustainable cities and communities | Make cities inclusive, safe, resilient and sustainable | Mellowcabs is a South African startup that manufactures and operates solar-powered electric pedicabs and improves first and last mile public transportation in urban areas (Mellowcabs, 2019) |
| **12** | Responsible consumption and production | Ensure sustainable consumption and production patterns | Fly Catcher Technologies is an Indian startup that develops storage units to convert organic waste into biogas for household use (DIPP, 2018) (also part of the Swachh Bharat Grand Challenge, see SDG 6) |
| **13** | Climate action | Take urgent action to combat climate change and its impacts | The Global Cleantech Innovation Program led by the United Nations Industrial Development Organization (UNIDO) and the Global Environment Facility (GEF) aims to strengthen the entrepreneurial innovation ecosystem in developing countries and supports demand-driven small and medium enterprises and startups (UNIDO and GEF, 2018). The Green Climate Fund (GCF) is planning a program to support climate technology incubators and accelerators in developing countries (Green Climate Fund, 2018; UNFCCC, 2018). |
| **14** | Life below water | Conserve and sustainably use the oceans, seas and marine resources | Evoware is an Indonesia-based startup that is developing edible seaweed-based packaging to replace plastic packaging that is a major contributor to marine pollution (New Plastics Economy, 2019) |
| **15** | Life on land | Sustainably manage forests, combat desertification, halt and reverse land degradation, halt biodiversity loss | Green Charcoal is a Uganda-based startup that replaces wood charcoal and firewood with agricultural waste (e.g. rice husk) (Green Charcoal Uganda, 2019). |
| **16** | Peace, justice and strong institutions | Promote just, peaceful and inclusive societies | UNDP Honduras developed a pilot 'Fab Lab' to 3D print hand prostheses for returning migrants and victims of violence who live with disabilities (UNDP, 2018a) |
| **17** | Partnerships for the goals | Revitalize the global partnership for sustainable development | The Climate Technology Center and Network (CTCN) of the UNFCCC consists of a worldwide network of organizations that support the development and transfer of climate technologies to developing countries (CTCN, 2016) |

The discourse on the types of government-led action needed for linking STI and SDGs has focused on setting up global, national, and subnational roadmaps that can help, among others, link STI policy with the 2030 Agenda and create enabling conditions to support a robust innovation system (IATT, 2018). Effective roadmaps need an evidence base that builds on specific experiences of developing countries in linking SDGs (and, in the past, related development goals) with policies for STI. In this aspect, as we show in the following sections 2.2 and 2.3, incubators have played a vital role.

[5] Goal 8 has specific targets that explicitly link to STI. For example, Target 8.2 aims for "economic productivity through diversification, technological upgrading and innovation…". Target 8.3 supports "…productive activities, decent job creation, entrepreneurship, creativity and innovation, and encourage the formalization and growth of micro-, small- and medium-sized enterprises…" (United Nations, 2015)

[6] Goal 9 has specific targets that explicitly link to STI. For example, Target 9.5 aims to "enhance scientific research, upgrade the technological capabilities of industrial sectors in all countries, in particular developing countries, including, by 2030, encouraging innovation and substantially increasing the number of research and development workers per 1 million people and public and private research and development spending." Target 9.B supports "domestic technology development, research and innovation in developing countries…" (United Nations, 2015)

### 2.2. Incubators for STI-based entrepreneurship

Incubators have been a vital element of STI policy (Aernoudt, 2004; Mian et al., 2016; Phan et al., 2005) in their role as intermediaries that link startups with networks of universities, investors, industry, government, etc. (Dutt et al., 2015; Peters et al., 2004). Incubators originated in the 1980s in the United States and have since been used worldwide (Akçomak, 2009; Allen and McCluskey, 1991). Estimates indicate nearly 7,000 incubator programs around the world of which a third focus on STI-based entrepreneurship (Mian et al., 2016). While various incubator configurations exist (such as accelerators and science parks), incubators are generally considered not-for-profit entities that receive varying levels of assistance from public funding bodies and provide long duration support for startups (up to five years) to help the conversion of individual early stage, high-risk ideas to marketable enterprises (Cohen, 2013; Dee et al., 2011; Hackett and Dilts, 2004; Mian et al., 2016). Incubators traditionally offer support services to startups including infrastructure (working space and associated basic physical infrastructure, workshops), finance, business capability (mentoring, training, consulting), and access to networks (Amezcua et al., 2013; Cohen, 2013; Dee et al., 2011; Dutt et al., 2015; Hackett and Dilts, 2004). Overall, incubators play a vital role in strengthening the ecosystem in which startups operate.

A rich body of literature on incubators in industrialized countries has shown that incubators have diverse types of primary goals and operational activities related to their incubatee startups, and these result in different types of innovation-related outcomes. For example, incubator goals could vary in the focus on broad STI or on the development and transfer of specific technologies, in the emphasis on economic development and employment generation, or in the linkages to universities or private sector, while outcomes could be linked to product-, process- or organizational- innovation (Barbero et al., 2014). In particular, STI-based incubators support the transfer of technology and help promote STI-based entrepreneurship. STI-based incubators with close linkages to universities or research centers (a) get access to knowledge-based assets (such as technically-trained students and faculty) (e.g., Jaffe et al., 1993; Rothaermel and Thursby, 2005a), (b) help in incubatee startup survival (Rothaermel and Thursby, 2005b), and (c) provide help for incubatee startups in developing networks (Lamine et al., 2016; McAdam et al., 2016; McAdam and McAdam, 2008). The linkages between STI-based incubators and universities not only have innovation-related outcomes but also help in implementing regional economic development goals as the localized clusters and regional networks formed promote entrepreneurial culture, information sharing, knowledge spillovers within and across firms and academia, and additional innovations (Saxenian, 1996).

While it is clear that incubator goals and activities dictate outcomes, there is no universal framework for assessing 'success' (Phan et al., 2005). Incubators have been subject to extensive scrutiny worldwide in terms of their formation and function, their performance outputs and outcomes, and on their linkages with public and private actors (Bergek and Norrman, 2008; Dee et al., 2011; Hackett and Dilts, 2004; Phan et al., 2005). However, there is no consensus on what defines success—measures of success could include survival, sales growth, employment growth, innovativeness of incubated firms, or meeting goals of the public sector (Akçomak, 2009). A meaningful analysis of incubator activities therefore requires an understanding of the context for entrepreneurship (Autio et al., 2014), the underlying goals under which a specific incubator was set up (Bergek and Norrman, 2008), and the different incubation strategies or incubation business models applied in relation to various goals (Clarysse et al., 2005; Grimaldi and Grandi, 2005).

Assessing the outcomes of existing incubators is critical for understanding the effectiveness of prevailing efforts and to justify future public spending in incubators for implementing SDGs. In the absence of a clear definition of incubator success, we use the term *effective* to describe incubators that have advanced SDGs (post 2015) or have advanced goals that can be mapped to the SDGs (pre 2015).

**2.3 Incubators for STI-based entrepreneurship in developing countries**

Mian et al., (2016) shows that the large body of literature on publicly-funded incubators is principally based in industrialized countries of Europe or the United States. Less analyses exist on incubators in developing countries because the lack of frameworks of assessment, the lack of clarity in metrics of success, and the stark differences in policy goals of promoting STI-based entrepreneurship compared to the goals in developed countries (see 2.1) exacerbate the challenges with assessment.

Research on incubation programs in developing countries remains limited to few countries (e.g., Akçomak, 2009; Chandra and Fealey, 2009; Chandra and Silva, 2012; Lalkaka, 2002; Scaramuzzi, 2002; Tang et al., 2013). For example, in Brazil, incubators emerged when the government interventions in innovation shifted in the mid-1980s from the former military regime's centralized large-technology projects to bottom-up innovation focusing on entrepreneurship at local and regional levels (Almeida, 2005; Chandra and Fealey, 2009; Etzkowitz et al., 2005). Brazilian incubators now have visible and active linkages to universities, industry, and government reflecting the 'triple-helix' model of synergies between these three stakeholders (Akçomak, 2009; Chandra and Fealey, 2009; Etzkowitz, 2002). In India, the government started to fund incubators since the mid-1980s as a means of developing STI-based entrepreneurial activity that would generate employment opportunities for a science and technology trained workforce (Lalkaka, 2002). In Chile, local governments collaborated with universities and industry to set up incubators since the early 1990s, paying particular attention to leveraging regional resources and to organizing risk capital and financing for early-stage startups as well as for incubators (Chandra and Silva, 2012). In contrast, incubators in China had a top-down mandate, with the government considering them as strategic avenues for technological advancement and economic development under China's transition to a high technology-driven market economy (Chandra and Fealey, 2009). The government enabled STI-based entrepreneurship by heavily funding and subsidizing a large number of high-technology incubators and their incubatees (giving special attention to the returning Chinese diaspora) (Akçomak, 2009; Chandra and Fealey, 2009).

The emphasis on STI-based startups for implementing SDGs (2.1) and the role that incubators play in enabling conditions for these startups (2.2 and 2.3) calls for a systematic understanding of historical incubator experiences in developing countries for maximizing the effectiveness of emerging policies and programs.

## 3. Case context and approach

The rest of the paper focuses on India and analyzes the role of incubators in enabling STI-based entrepreneurship under evolving STI policy priorities.

### 3.1. Science technology and innovation in India

India is one of the world's fastest growing economies but its multiple market failures mean that it does poorly in many societal and environmental aspects, e.g. human development, income inequality, and greenhouse gas emissions (Alvaredo et al., 2018; UNDP, 2018b). Balancing economic growth with sustainable development challenges is thus not only central to policymaking in India, it reflects the challenges many developing countries currently face or can expect to face as they grow.

Specifically, India is a suitable candidate for a case study on the role of STI-based incubators and startups and their potential for implementing SDGs because of three reasons. *One*, India has had over three decades of experience in linking STI with development goals, well before they were formalized in the SDGs in 2015. While countries like Brazil and China also have had multi-decade experiences, the context in which incubators emerged and STI policy evolved in India is shared by many other developing and emerging economies—e.g. in its experience with economic liberalization reforms, emergence of multinational companies and their R&D centers, the return of IT-trained diaspora interested in exploiting new entrepreneurial opportunities, and in its efforts to match global trends in emphasizing new and emerging models of innovation through startups.

*Two*, India's low score in various indicators of STI (Figure 1) and low levels of R&D investments, research personnel since the 1990s, and of patenting activity (Figure 2) are comparable to the situation in many other developing countries. Apart from a few leading institutions, the record of India's technical higher education is rather spotty – only up to half of the engineering graduates are seen as employable (CII et al., 2018; Tandon, 2017). A large majority of small business that claim to engage in innovative activity do so on the basis of their use of new machines rather than their development of new STI-based processes or products (NSTMIS - DST GoI, 2014). While India has always been an entrepreneurial country known for frugal, flexible innovations (e.g., Radjou et al., 2012) and some widely cited reports suggest that its startup ecosystem follows that of the US or China (NASSCOM, 2015), not all entrepreneurial activities are associated with STI-based startups[7]. While such startups generate revenues, create employment, and are important for supporting economic growth, they represent business model innovations rather than STI. This means that India's low score in STI indicators, its weaknesses in higher education, as well as its poor outcomes in terms of STI-based startups are reflective of the situation in other developing countries, especially when compared to China where many of these metrics are significantly higher.

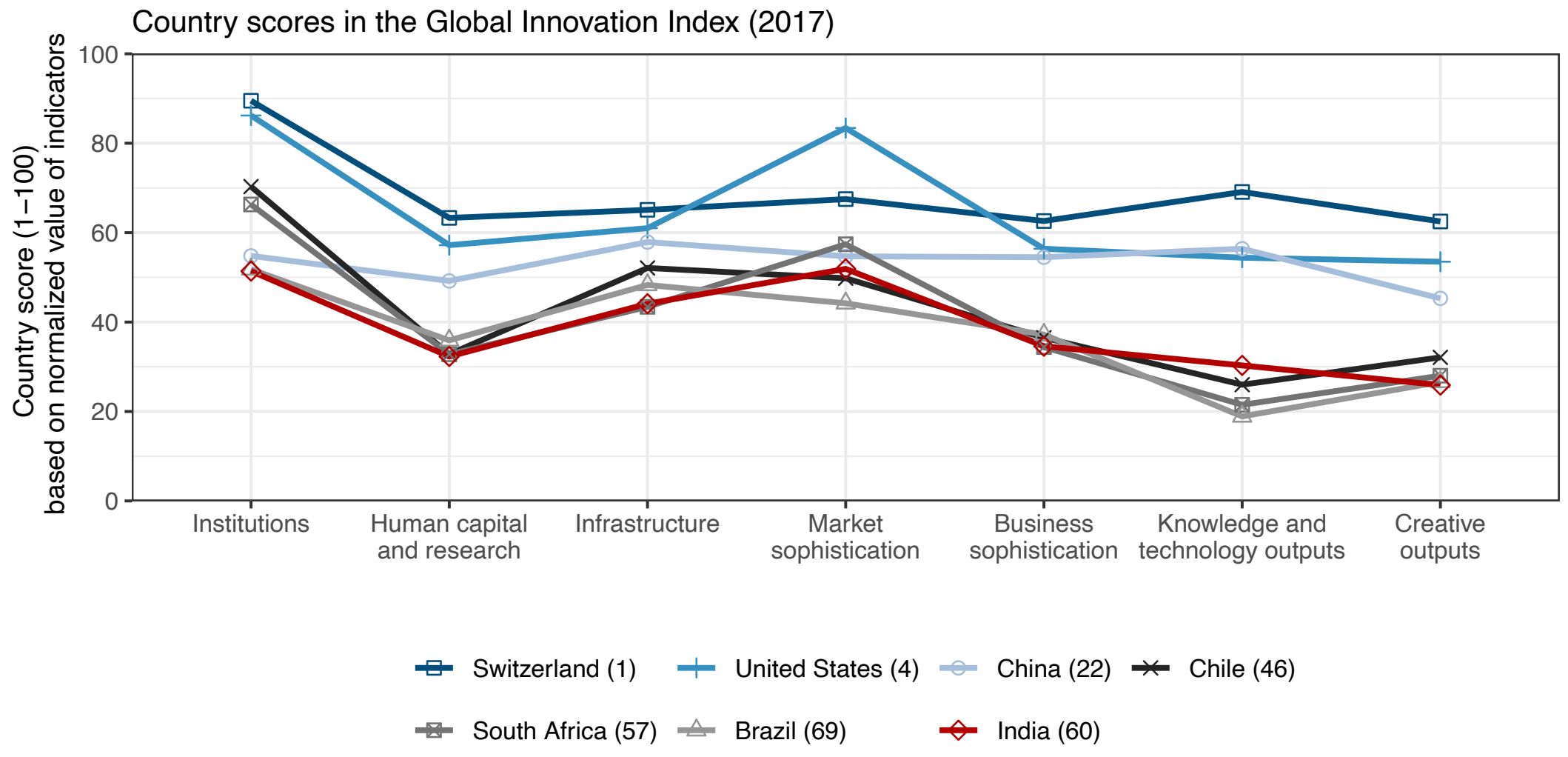

FIGURE 1: Selected country scores in the Global Innovation Index (2017) show that India scores low in various indicators of innovation compared with other major industrialized and developing countries. Furthermore, all developing countries demonstrate weaknesses in human capital, education, and research as well as knowledge,

[7] Despite the growing number of successful startups and entrepreneurs in India, not all entrepreneurs innovate in science and technology. Most commercially-successful Indian startups of the mid-2010s—for example, Naukri.com, Flipkart, Ola, Snapdeal, Zomato—have used established business ideas with proven international success and adapted them in the Indian market (see Raghavan, 2016).

technology, and creative outputs. Brackets next to countries show country ranking. Source: Cornell University et al., 2017.

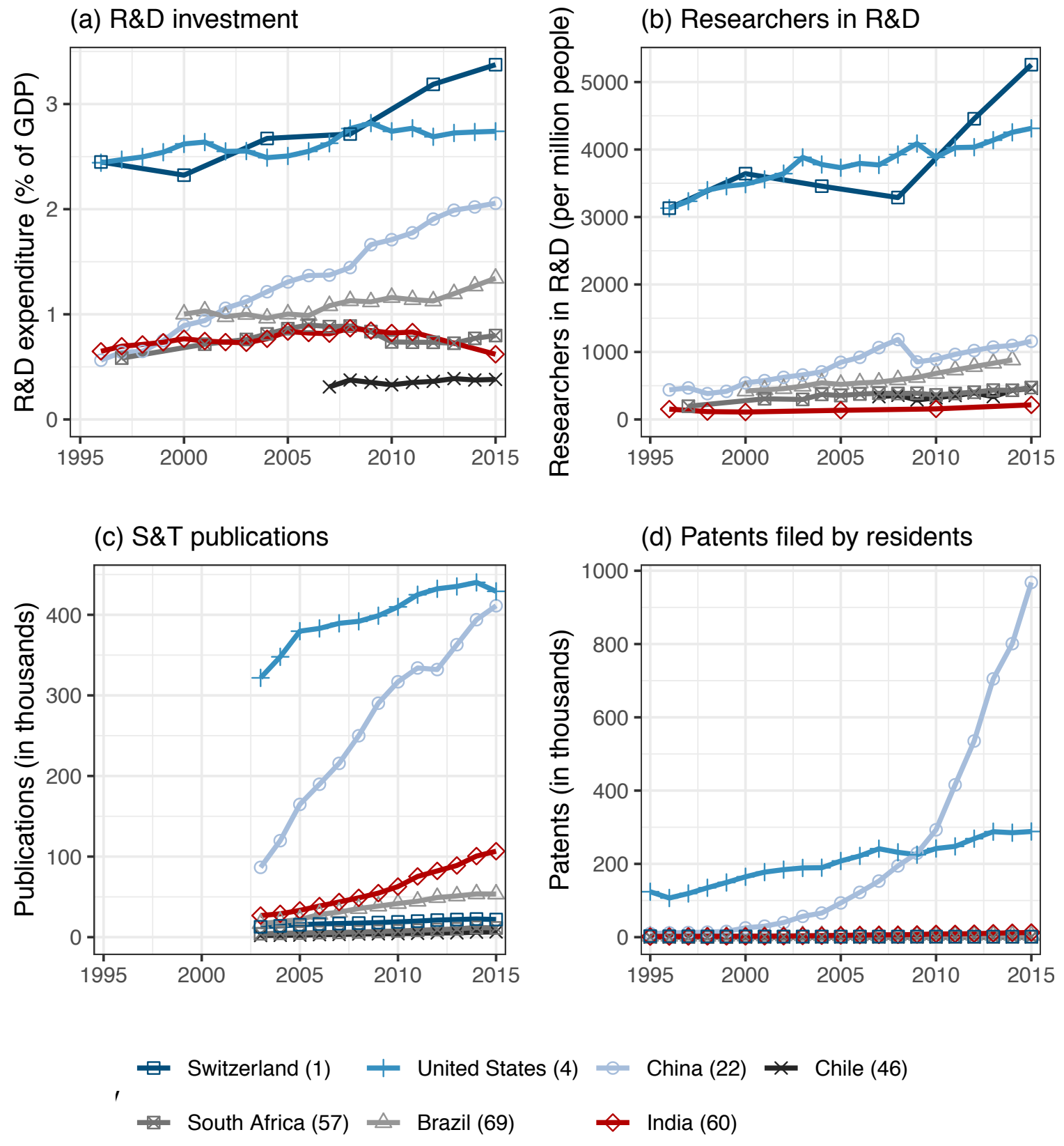

FIGURE 2: Various indicators of science, technology, and innovation illustrate India's continuously low inputs (e.g., (a) R&D investment), activities (e.g., (b) number of researchers) and outputs (e.g., (c) publications and (d) patents). Low R&D investments contribute to low number of researchers. And, even though science and technology publications have increased over time, they do not indicate an increase in STI given the small number of patents filed. Overall, India's efforts are low compared to major economies such as China where the emphasis on STI rapidly took off in the past decade. Source: World Bank, 2017

*Three*, India's STI-related activities (including but not limited to R&D) have specifically linked to economic and sustainable development-related goals and incubators have been particularly important in the implementation of that effort. This engagement has evolved from the post-Independence approaches since the early 1950s in more centralized R&D[8] and technology transfer from large public-sector agencies, governmental laboratories, and large industries and private firms, to supporting specific sectors such as biotechnology or information technology (IT), and more recently to a decentralized emphasis on STI-based entrepreneurial startups since the 2000s. The importance of incubators is evident in the establishment of the National Science and Technology Development Board that set up incubator-like entities since 1985.

### 3.2. Approach

We used process tracing (Bennett and Checkel, 2015; Collier, 2011) to assess the linkages STI policy and related publicly-funded incubators in India with societal and developmental goals (these goals existed for multiple decades but were organized around the SDGs after 2015). We used extensive archival research of government documents and semi-structured interviews (Appendix Table A1) with a variety of stakeholders for each of the following steps. First, to establish the links between STI policy and societal and developmental goals, we analyzed the evolution of public policy goals related to STI-based entrepreneurship and incubators over time. Next, to understand how policies for incubators were actually implemented by different actors in the innovation system, we analyzed the activities of a complex network of public agencies who funded incubators. Finally, to infer how individual incubators were able to actually implement the policy goals identified earlier, we identified

[8] These public institutions include (i) laboratories, e.g. the Council for Scientific and Industrial Research CSIR, (ii) large scientific agencies e.g., Department of Atomic Energy (DAE), Indian Space Research Organization (ISRO), (iii) public-sector enterprises e.g., Bharat Heavy Electricals Limited (BHEL), Hindustan Machine Tools (HMT), Indian Oil Corporation Limited (IOCL), and (iv) technical higher education institutions e.g., Indian Institutes of Technology IITs

six incubators (with the help of analysis and interviews of the previous two steps) that indicated effectiveness in implementation of societal and developmental goals using STI-based entrepreneurship.

Because of the absence of a clear definition or metrics of incubator ‘success’, the six case studies that we identified as effective in meeting societal and developmental goals were based on inference from our analysis complemented with inputs from expert interviewees (Table 2). Our approach on the validating the study and its findings with multi-step interviews was driven by the lack of data on public funding for incubator-related activities—both in terms of how (and how much) funds were allocated and what have been the outcomes of public-funding or policy interventions. The lack of systematic data made it difficult to quantitatively assess over time all incubators, their interactions with government and other actors, or their changing goals, priorities, outputs, outcomes etc. We partially addressed issues related to the lack of data by developing our own database from 1985-2014 of all publicly funded incubators that included public entities that received full or partial public-funding to support STI-based entrepreneurship—this helped us identify an initial set of incubators which we then refined with the help of interviews.

While our six incubators are by no means an exhaustive representation of incubation activities in India, they represent incubators with a spectrum of goals, locations, and partners that reflect the various operating conditions that incubators in developing countries may face—i.e., in locations with access to different types of resources (e.g., location in metropolitan Tier I city[9] with extensive financial and industrial resources and networks vs. location in smaller cities or towns with limited resources or networks), different sector-related priorities (e.g., biotechnology vs. information technology), and different incubator partners (e.g., engineering university vs. business school). For each individual incubator, we conducted semi-structured interviews (see Appendix table A2 for a list of questions) with the goal of understanding the evolution of incubator activities that had enabled implementation of societal and developmental goals that are now organized around SDGs.

TABLE 2: Overview of incubators analyzed in this paper

| **Incubator** | **Host or Partner** | **Host type** | **Location** | **Focus** | **Type of location** |
|---|---|---|---|---|---|
| Center for Innovation Incubation and Entrepreneurship (CIIE) | Indian Institute of Management, Ahmedabad (IIMA) | Business School | Ahmedabad[1] | Information technology, energy, water, agriculture, IoT, fintech, entrepreneurship | Tier-1 city |
| Centre for Cellular and Molecular Platforms (C-CAMP) | Bangalore Biotech Cluster | Central Government Research Lab | Bangalore[1] | Biotechnology, late-stage science | Tier-1 city |
| Incubator at IKP Knowledge Park (IKP) | IKP Foundation | Foundation | Hyderabad[1] | Biotechnology, hardware products, healthcare, medical devices | Tier-1 city |
| Society for Innovation & Entrepreneurship (SINE) | Indian Institute of Technology, Bombay (IITB) | Engineering College | Mumbai[1] | Information technology, intellectual property-based ventures, student entrepreneurship | Tier-1 city |
| Startup Village (SV) | MobME, Government of Kerala | Private Company | Kochi[2] | Information technology, acceleration, electronics | Tier-2 city |
| Technology Business Incubator – Kongu Engineering College (TBI-KEC) | Kongu Engineering College (KEC) | Engineering College | Perundurai, Erode[3] | Technology-led entrepreneurship | Tier-3 city |

[9] Tier I cities are Delhi NCR, Hyderabad, Bangalore, Mumbai, Kolkata, Ahmedabad, Pune

## 4. Results and discussion

An assessment of the historical evolution of STI policy provides insights into *why* incubators were set up (4.1), what various stakeholders involved in incubator activities did and what were the outcomes (in 4.2), and how they were able to implement SDGs (4.3).

### 4.1 Evolution of public policy for STI-based entrepreneurship

Our analysis of the evolution of public policy related to STI-based entrepreneurship (Table 3) finds that incubators have been central to implementing the broader STI-based policy goals from the 1980s to the mid-2010s. In that, even prior to 2014, incubators have been vital to implementing the policy goals that can now be mapped to Sustainable Development Goals 8 and 9—most prominently in supporting entrepreneurship and economic growth and supporting information technology (IT) and biotechnology industries.

TABLE 3: Evolution of public policy goals for STI-based entrepreneurship and incubators in India. Source: Authors' compilation from Five Year Plan (FYP) reports and other sources

| Period | Announced plans and policies | Broader policy goals for STI |
|---|---|---|
| 1980 – 1984 | • 1982: Department of Science and Technology (DST) sets up the National Science and Technology Entrepreneurship Development Board (NSTEDB)<br>• 1984: NSTEDB starts the Science and Technology Entrepreneurs Park (STEP) program | |
| 1985 – 1989 | • 1986: Government of India sets up the Department of Biotechnology (DBT)<br>• 1987: DST sets up three pilot incubators with United Nations Fund for Science and Technology | • Generating employment for science and technology-trained workforce |
| 1990 – 1991 | • 1991: Government of India engages in country-wide economic liberalization reforms | |
| 1992 – 1996 | | • Generating employment for science and technology-trained workforce<br>• Training entrepreneurs for biotechnology<br>• Commercializing indigenous technology |
| 1997 – 2001 | • 2001: NSTEDB sets up the Technology Business Incubators (TBI) program | • Regional development<br>• Training entrepreneurs |
| 2002 – 2006 | • 2004: Indian Step and Business Incubator Association (ISBA) created<br>• 2004: DST sets up the Technology Development Board (TDB) seed fund for financial support of startups | • Establishing interfaces between academia, R&D, and industry<br>• Training rural populations in IT to encourage entrepreneurship<br>• Supporting grassroots innovation<br>• Developing biotechnology through creating a venture capital fund, commercializing technologies, creating incubators and science parks |
| 2007 – 2011 | • 2007: DST incubators and incubatee startups are exempt from paying service tax<br>• 2008: Department of Electronics and Information Technology (DeitY) launches the Technology Incubation and Development of Entrepreneurs (TIDE) scheme for supporting electronics, IT startups | • Supporting STI-based entrepreneurship<br>• Fostering academia-industry linkages;<br>• Commercializing technology developed at universities using incubators<br>• Supporting startups financially by facilitating venture funding and tax incentives<br>• Encouraging entrepreneurs through flexible salaries, startup grants<br>• Supporting biotechnology through incubators, parks, and clusters, and through public-private partnerships |
| 2012 – 2016 | • 2012: Small and Medium Enterprises (SME) exchange / trading platform launched<br>• 2012: DBT launches Biotechnology Industry Research Assistance Council (BIRAC)<br>• 2012: BIRAC initiates the Biotechnology Ignition Grant (BIG) program<br>• 2013: DST, MoMSME incubators qualify for Corporate Social Responsibility spending<br>• 2013: Startups list on SME exchange<br>• 2013: DST announces a Science, Technology, and Innovation Policy<br>• 2014: Government of India revises bankruptcy laws<br>• 2014: Government of India encourages venture capital, angel investors<br>• 2015: State governments' launch startup policies<br>• 2015: Atal Innovation Mission<br>• 2015: National Policy on Skill Development and Entrepreneurship<br>• 2016: Startup India Action Plan | • Building an inclusive innovation system across sectors for entrepreneurship, growth<br>• Supporting biotechnology innovation with incubators, parks, and clusters for technology transfer and management; new funding schemes from public-private partnerships and with BIRAC |

For three decades, incubators and startups were central to STI policy, first in their perceived ability to generate employment through supporting new enterprises (1980s), then in building academia-industry and science-technology linkages and supporting technology transfer (early 2000s), and then in the specific endeavor to promote startups (2010s onwards). In the early 1980s, to encourage STI, the government engaged in incubator-building programs as it set up the administrating body, the National Science and Technology Entrepreneurship Development Board (NSTEDB) in 1982, that continues to administer publicly-funded incubators to date. The first incubators were set up to generate employment through Science and Technology Entrepreneurs' Parks (STEPs), established prior to the economic liberalization reforms of the early 1990s when economic growth was particularly low and the innovation system particularly weak. The general paucity of innovation and entrepreneurial activity attracted regular firms interested in the basic infrastructures that STEPs offered (e.g., space and an improved supply of water and electricity) rather than STI-based startups (Mittal, 2015). These regular firms did not graduate as incubatees even after several years, thus contributing to financial challenges and the systematic failure of the STEP program (DST, 2014)[10].

By the early 2000s, a new program for Technology Business Incubators (or TBI) focusing on STI-based startups built on learning from past experiences and shortcomings of the STEP program as well as from other failed pilot incubators[11]. Meanwhile, STI capabilities had grown because of the pairing of liberalization reforms of the early 1990s that led to greater availability of technology (as import tariffs lowered) with the increase in IT-trained talent (through reverse brain drain after the global dot-com bubble of the late 1990s). Incubators became a core part of STI activities, as reflected in the creation of an incubator association or Indian STEPs and Business Incubators Association (ISBA) designed to foster networks and to share best practices (Ministry of Science and Technology, 2004).

In the late 2000s and early 2010s, the public policy goals of strengthening linkages with industry and encouraging STI-based entrepreneurial activities generated new incentives channeled through incubators. Publicly funded incubators and their tenant incubatee startups became exempt from paying service tax in 2007. In 2013, the Corporate Social Responsibility (or CSR) program included

[10] Of the 16 STEPs that were established between 1984 and 1995, only 6 demonstrated results or financial sustainability by 2001.
[11] Incubators had been established in 1987-1990 by the United Nations Fund for Science and Technology (Lalkaka, 2002)

spending on publicly funded incubators in its scope of activities related to social goods—the CSR program required corporate companies with high net worth and profits to spend two percent of their profits on social issues (*Companies Act*, 2013).

With this rich experience in setting up incubators, policies and programs enacted since the mid-2010s raised their ambition in linking STI with societal goals (and mapped them to the SDGs) with the support of STI-based incubators. National- and state-level policies and programs emerged that specifically targeted the creation of new enterprises and innovation (e.g., Make in India, Startup India, Atal Innovation Mission, the National Entrepreneurship Policy, and the National Policy on Skill Development and Entrepreneurship (UN India, 2019a, 2019b)) and the promotion of industry development, especially for biotechnology. Startup India, now mapped to SDG 9 primarily targeted practical barriers to innovation through: (i) easing of complex, lengthy regulatory processes for startups, (ii) providing high-risk funding and tax incentives to startups (with a total budget of INR 100 billion to be distributed by 2020), and (iii) promoting industry-academia linkages including through 70 new incubators, startup centers, and research parks (Ministry of Commerce and Industry, Government of India, 2016). The Atal Innovation Mission now mapped to SDG 8 aimed to address social and economic development issues through STI by: (i) building the capacity to innovate in middle- and high-school students through 500 new maker-spaces known as Atal Tinkering Labs, (ii) creating 100 new sector- or technology-specific incubators, and (iii) extending support for existing incubators. While these policies and programs collectively targeted an intensification of incubator activity they were built in an absence of lessons learnt from past experiences.

## 4.2 Implementing policies: characterizing publicly-funded incubator activity

A multitude of public agencies were involved in implementing the various STI policy goals related to setting up incubators. The activities and characteristics of these agencies—i.e. *what* they did—influenced the outcomes of incubator programs managed by these agencies as well as the outcomes of individual incubators that they funded.

Our assessment shows that government ministries and departments implemented policies in support of STI-based entrepreneurship and related SDGs in two ways. *First*, government departments—most prominently Department of Science and Technology (DST) through the National Science and Technology Entrepreneurship Development Board (NSTEDB), and the Department of Biotechnology (DBT) through the Biotechnology Industry Research Assistance Council (BIRAC)—engaged in developing incubators. Several other governmental departments indirectly supported existing incubators to enable STI-based entrepreneurship in order to promote a particular sector (e.g., The Electronics and IT Department (DeitY) for IT) or to advance a particular agenda (e.g., promotion of small and medium enterprises through the Ministry of Micro Small and Medium Enterprises (MoMSME)) (Upadhyay et al., 2010).[12] *Second*, government departments supported startups and entrepreneurial activity by providing access to government funding for those startups that were located in publicly-funded incubators (rather than private accelerators which are not part of this study). Startup incubatees had access to governmental financing designed for startup survival until the technology demonstration stage before STI-based startups were investment-ready.

[12] The Department of Scientific and Industrial Research (DSIR) with its mandate of advancing industry-centric research and innovation ran a grants funding program for incubators—Promoting Innovations in Individuals, Startups and MSMEs, or PRISM. The Electronics and IT Department (DeitY) focused on electronics and IT-related industries and digital services by offering financial support to existing incubators through the Technology Incubation and Development of Entrepreneurs (TIDE) scheme. The Ministry of Micro Small and Medium Enterprises (MoMSME) collaborated with host institutes to designate incubator-like entities that encouraged early-stage ideas in a range of sectors (biotechnology, nanotechnology, fruit processing, ceramics, surgical instruments, etc.) (MoMSME, 2010). Others include the Ministry of New and Renewable Energy (MNRE), the Department of Industrial Policy & Promotion (DIPP) of the Ministry of Commerce and Industry, and several state departments.

The approaches used differed between departments. The DST promoted STI-based startup creation across all sectors through the NSTEDB. NSTEDB solicited applications for setting up incubators, and approved those that were set up with a partner 'host' institute (i.e., a university or R&D center) among other criteria (NSTEDB, 2012a). Once approved, the NSTEDB provided initial financial support for five years for setting up and managing the incubator. In addition, DST provided funding for startups located in its incubators through the Technology Development Board (TDB) and the Seed Support System that provided financial assistance (through debt, equity share, or a share of royalties) to technology-focused startups physically located in government-approved incubators (NSTEDB, 2012b)[13].

The DBT supported innovation through BIRAC, a public-sector enterprise for facilitating creation of biotechnology-based startups and converting research into products. BIRAC implemented the Bio-Incubators Support Scheme that aimed to create new incubators and to strengthen established, proven incubators. In addition, BIRAC provided financial support for entrepreneurial activity through several funding mechanisms—e.g., the Biotechnology Ignition Grant for developing early stage proof-of-concepts and the Small Business Innovation Research Initiative for growth. The

[13] The NSTEDB provided over INR 2 billion in funding for incubators, evolving from around INR 2 million for each incubator in the late 1980s, to an average of INR 30 million by 2015 (Gupta, 2015).

Biotechnology Ignition Grant required applicants to be an incubatee in an eligible DBT incubator or to have a registered company with a functional R&D laboratory. In general, the DBT focused on the specific needs of innovation the life sciences industry—e.g., (i) long gestation period for startups (about five years) (ii) high capital intensity of technologies; and (iii) need for highly-skilled workforce to operate technical equipment.

The outcomes of policy goals (4.1) and implementation activities related to STI-based entrepreneurship resulted in at least 140 incubators that received public funding through different central government agencies between 1985 and 2014. Of these, at least 86 Technology Business Incubators (TBI) and Science and Technology Entrepreneurs Parks (STEP) received funding from the NSTEDB (DST, 2014), making it (and the DST) the single largest supporter of incubators. Notably, we found the data on the number of incubators, their activities and funding sources, and their outcomes to be highly inconsistent (Table 4), and we therefore point out that the estimate of at least 140 publicly-funded incubators is based on our own in-depth assessment and data development.

TABLE 4: Estimates of number of incubators in India from the DST (DST, 2014, 2009), governmental policy research and planning (NITI Aayog, 2015) and a leading IT industry association (NASSCOM, 2015) are inconsistent. We found 146 publicly funded incubators (including those with MoMSME) with cumulative support for at least 2000 startups from 1985 through 2014.

| **Description** | **DST 2009** | **DST 2014** | **Niti Aayog 2015** | **Nasscom 2016** | **Authors' Calculations** |
|---|---|---|---|---|---|
| Number of incubators | 36 reported (full or partial data on publicly funded incubators) | 54 reported (publicly funded incubators) | 120 | 140+ (incubators and accelerators; public and private) | 146 government supported incubators |
| Employment generated | 13,400+ | 32,000+ | 40,000 | 100,000 (in all startups) | - |
| Startups | 1170+ (incubated)<br>486+ (graduated) | 2000+ (incubated)<br>950 +(graduated) | 800+ (graduated since 1982)<br>500 supported annually | 4,750+ | 2000+ (supported) |

Our analysis of the existing publicly-funded incubators in India reveals several shared characteristics in how *most* incubators operate. The following (and Figure 3) discusses the full set of these incubators (including our six case studies), their locations, and how they tap into different public-funding sources, illustrating their shared characteristics and highlighting some differences. *First*, fitting the traditional description of incubators (2.1), STI-based incubators operate as a not-for-profit entity (as a registered society), or as a company that is required to reuse profits or income and cannot provide any dividends to shareholders (a Section-25 company). *Second*, most incubators are set up with a host academic institute or a R&D laboratory with the expectation that the host provides technology infrastructures needed for STI along with space and other basic facilities. 84 percent of NSTEDB incubators have such a host partner (DST, 2014). *Third*, in addition to indirectly supporting funding in their incubatee startups by attempting to strengthen networks with private investors, angel investors, venture capital, etc., a key role of incubators in India is to facilitate direct financing for startup incubatees from public funding channels that can only be offered to startups located in these government-approved incubators (See Figure 3). *Fourth*, incubators receive five years of financial support (those supported by the DST), after which they are expected to sustain their own business (DST, 2014). Other government departments also have some provisions for incubators to support their expenses (e.g. for SINE). While the DST remains the most significant source for most STI incubators, the DBT provides substantial support specifically for biotechnology and life sciences incubators (e.g. C-Camp) (Figure 3). But for most incubators, public support is not enough and they may generate revenues by renting out infrastructures or by providing services to tenant enterprises rather than focused engagement in STI entrepreneurship. And *fifth*, most incubators are in clusters around metropolitan Tier 1 cities that have strong industrial presence (Figure 3). These incubators (e.g. CIIE, IKP, SINE) are often able to tap into a larger number of public funding resources to support themselves as well as their startups, likely due to both higher quality of startups and a relatively stronger innovation system in the region. When located outside of these clusters (e.g. the large number of MoMSME-only supported incubators and the exception that is KEC), incubators specifically aim to contribute to regional development and support small businesses that generate STI-based entrepreneurship.

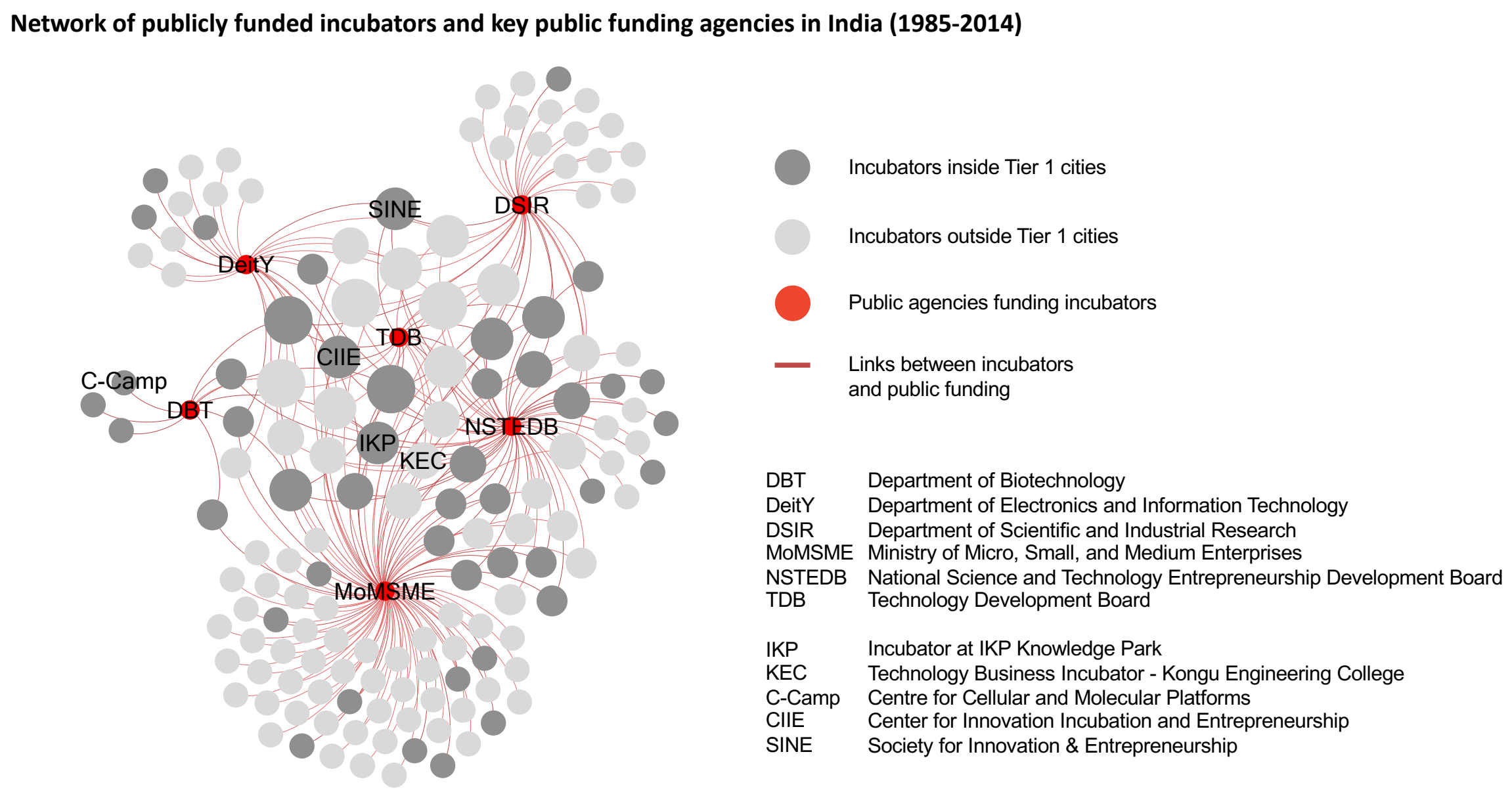

FIGURE 3: The network of publicly funded incubators in India and their funding sources shows that most incubators are funded by multiple sources (including those analyzed in this paper); these are represented by larger size circles. Funding sources after 2014 (such as Atal Innovation Mission) are not included.

But despite the large presence of incubators in urban metropolitan clusters with strong industry presence, we found that the actual linkages of most incubators with industry had been weak (exceptions include DBT incubators (Aggarwal and Chawla, 2013) and some DST incubators).

Public policy interventions on strengthening these linkages had been ineffective. For example, the eligibility of publicly funded incubators for Corporate Social Responsibility spending was inconsequential for strengthening incubator-industry linkages as corporate firms preferred to directly finance more popular social values in support of government initiatives (e.g. Swachh Bharat Mission related to clean water and sanitation) rather than address these goals through riskier support for STI and entrepreneurship in incubators. Furthermore, when existent, most incubator relationships with industry were not through Indian public sector units or domestic firms but through large multinational corporates interested in developing domestic technologies for strengthening their supply chain networks.

Overall, while some incubator characteristics encompass the activities of a traditionally defined incubator (in section 2.1) that is not specifically linked to STI or to developing countries, there are aspects of both the 'incubation system' in India, and of individual incubators that stand out as extending beyond traditional activities. The incubators that were able to implement policy priorities and meet the societal and developmental goals linked to SDGs were able to do so because of their individual characteristics rather than broader program, industry, or location support.

### 4.3 How incubators implemented STI-based entrepreneurship for SDGs

Our detailed analysis of six incubators identifies common features that show *how* incubators were effective in specifically enabling STI-based entrepreneurship and implementing clear goals that could eventually be mapped to the SDGs. We found that these effective incubators primarily implemented Goal 8 and Goal 9 and in the process extended their activities well beyond traditional incubator functions (described in section 2.1). We summarize these extended activities in Table 5 and discuss them in the following.

TABLE 5: Incubator activities in our case studies that enabled STI-based entrepreneurship for SDGs

| **Incubator** | **Activities linked to promoting STI** | **Traditional and extended activities linked to promoting entrepreneurship** | | | | | **Illustrative list of Sustainable Development Goals mapped to incubators' goals** |
|---|---|---|---|---|---|---|---|
| | | **Attracting innovators through capacity building efforts** | **Addressing unmet market demands** | **Providing financing for startups** | **Building networks for startups** | **Managing incubators effectively** | |
| Center for Innovation Incubation and Entrepreneurship (CIIE) | • Focusing on connecting technology-based products with markets<br>• Building depth in STI-oriented sectors through market research, and networks (e.g., in energy, water, agriculture, internet of things, fintech) | • Organizing a national business plan competition (attracting and mentoring over 500 entrepreneurs in 2016)<br>• Using networks and resources of host business school<br>• Engaging in capacity building activities in business school as well as in the country | • Analyzing gaps in the innovation ecosystem<br>• Organizing self-reflection exercises to ensure gaps are addressed Incubating firms in sectors that meet societal needs | • Investing directly in startups through their own seed fund<br>• Facilitating DST financing | • Utilizing business school networks<br>• Offering virtual mentoring / incubation | • Generating revenues from seed fund, management fee<br>• Building effective partnerships with private sector<br>• Improving operations with internal evaluation<br>• Assessing gaps and conducting self-assessment and exercises | Goal 2 (agriculture), Goal 3 (health), Goal 7 (energy), Goal 8 (economic growth) |
| Startup Village (SV) | • Collaborating with DST for developing a regional innovation system focused on IT | • Partnering with local government<br>• Engaging in capacity building | • Fostering an entrepreneurial culture and generating entrepreneurial opportunities in a region with very little entrepreneurship | • Investing directly in startups<br>• Facilitating DST financing | • Organizing events and providing exposure to successes such as the Silicon Valley | • Generating revenues from seed fund, management fee, private sector | Goal 8 (economic growth) |
| Society for Innovation & Entrepreneurship (SINE) | • Commercializing STI-based research from partner engineering institute | • Using networks of and resources of host engineering school | • Addressing lack of product innovation by focusing on intellectual property and encouraging product oriented companies | • Facilitating DST financing | • Setting up a mentor pool with mentor for each incubatee | • Allowing startups to pay using a varied mix of rent, equity and revenue share and changing practices according to market conditions | Goal 9 (research and development) |
| Technology Business Incubator - | • Building regional innovation system in IT and | • Engaging in capacity building activities in partner engineering | • Incubating firms focusing on | • Facilitating DST financing | • Partnering with | • Maximizing revenue | Goal 8 (economic growth), |
| Kongu Engineering College (TBI-KEC) | related engineering technologies | school through entrepreneurship awareness camp for undergraduate students<br>• Partnering with local industry | product-based IT | | local industry association | through lean operations<br>• Workshops and skills training for local industry | Goal 9 (IT industry) |
| Centre for Cellular and Molecular Platforms (C-CAMP) | • Developing new technological platforms for biotechnology academia and industry<br>• Commercializing early stage biotechnology research<br>• Fostering linkages by co-housing academic scientists and startups | • Engaging with research and industry cluster<br>• Managing and distributing prestigious grants<br>• Housing startups and individual innovators selected through national competition | • Incubating firms focusing on life science and biotech needs<br>• translating high risk, yet promising research towards industry-ready technology | • Facilitating DST, BIRAC financing | • Supporting networks by proving a mentor on board, organizing forums and events | • Generating revenues from technology platform licensing fee<br>• Improving operations with internal evaluation | Goal 9 (biotechnology and life sciences industry) |
| Incubator at IKP Knowledge Park (IKP) | • Developing a regional innovation system around biotechnology by building a science park followed by an incubator | • Engaging with research and industry cluster<br>• Grand Challenges Exploration to attract bold new ideas in partnership with the Bill & Melinda Gates Foundation (BMGF) in 2011 and USAID in 2012<br>• Managing and distributing prestigious grants | • Incubating firms focusing on life science and biotech needs<br>• Creating new organizational structures (such as makerspaces) to address lack of product startups in India | • Investing directly in startups through the India Innovation Fund with partners<br>• Facilitating DST, BIRAC financing | • Offering virtual mentoring<br>• Regional innovation system mapping | • Generating revenues from seed fund, management fee, grant management<br>• Improving operations with internal evaluation | Goal 9 (biotechnology and life sciences industry) |

### 4.3.1 Identifying and attracting innovators

Public policies for STI-based entrepreneurship had historically focused on the creation of incubators but had put limited emphasis on building capacity for STI-based entrepreneurship (see Table 3). This led to a gap between incubators' high demand for quality ideas and innovators and the low supply of innovative, cutting-edge, technical ideas that were a consequence of insufficient talent, weak STI capacity in Indian academic institutions, and the relatively low understanding of markets and sectors relevant for sustainable and economic development goals (described in Section 3).

Incubators that were able to effectively implement goals related to the SDGs directly addressed this gap by engaging in capacity development activities or by benefitting from the presence of well-developed human capacity[14].

Three incubators CIIE, Startup Village SV, and TBI-KEC used the resources of their host partners while engaging in capacity building activities to attract innovators and develop ideas. CIIE's association with a business school of 'National Importance' (Indian Institution of Management Ahmedabad) offered it access to skilled students and networks related to businesses and market-needs in and beyond STI. CIIE further developed this resource by focusing on innovation-specific capacity building activities. More concretely, for the business school in the attached host institution, CIIE offered internship opportunities at the incubator, fellowships for student-entrepreneurs, prototype grants to offset concerns on education loans, and specialized courses (e.g., on mock fund management, technology and design). This provided a safe space for students to experiment with entrepreneurship and helped create a pool of entrepreneurs and early stage employees where none existed before. Outside the business school, CIIE attracted nation-wide ideas by hosting competitive programs (e.g., Power of Ideas) or by managing sector-based accelerator programs that addressed market-specific needs for societal and developmental goals. This helped nurture good ideas which might have been not funded had CIIE not existed. Similarly, Startup Village SV engaged with the state government and helped make innovation a priority for the state (through the Kerala Innovation Policy). Furthermore, SV prioritized capacity-building in its region to ensure the supply of innovators in the long-term, for example by advocating for and helping implement a program providing open-source electronics prototyping kits to selected school students to encourage experimentation and building innovative products. Finally, TBI-KEC in a regional engineering college away from a major metropolitan city managed to attract innovators and market-driven ideas despite limited resources. TBI-KEC countered its modest geographically-linked STI resources by engaging in capacity building activities—for example, offering training in entrepreneurship and organizing workshops for students and researchers to strengthen skills in specific IT areas (e.g., Very-Large-Scale-Integration (VLSI) design, embedded technologies). TBI-KEC complemented capacity building activities to attract innovators by engaging with local industry associations in the closest city (i.e., Coimbatore) and participating in industry-specific trade fairs.

[14] In India, while the government recognizes the need for building capacity among students and academic researchers, existing government-led efforts have been largely insufficient for reaching the number of students and researchers necessary for enabling transformational change. For example, existing plans under the Atal Innovation Mission to build 500 Tinkering Labs stand to benefit less than 0.7% of 72,000 senior secondary schools. Similarly, plans to build 300 university-affiliated incubators will benefit less than 40% of over 770 universities. Furthermore, faculty and researchers in most universities lack incentives to generate market-driven ideas—faculty hiring and promotion has been based on guidelines set by the University Grants Commission (UGC) that prioritized degrees and publications (UGC, 2016, 2013). In 2016, UGC guidelines for evaluation or promotion focused on publications and included patents, but did not specify entrepreneurship or startups as favorable metrics for faculty evaluation and promotion. Also, UGC rules for 'study leave' mainly supported research projects only and did not allow full- or partial- employment with any organization during the study leave period, possibly due to potential conflict of interest.

The three other incubators with a well-defined emphasis on STI activities relied on the resources of their partners or on their location to get access to good ideas and human capacity. In the case of SINE, its association with an engineering 'Institute of National Importance' delivered access to skilled engineering students, researchers, and alumni networks. The life sciences incubators' (C-CAMP and IKP) were built in physical proximity to biotechnology research and industry clusters that ensured access to scientific innovation and entrepreneurs. The access to knowledge and incubatees extended beyond their physical locations as these incubators managed or distributed several early-stage startup grants (e.g., from Biotechnology Ignition Grant, Bill and Melinda Gates Foundation) that attracted incubatees from the country.

### 4.3.2 Addressing unmet market demands

The policy goals for STI-based incubators in India (Table 3) targeted supply-side efforts—i.e., supporting technology push (e.g. for IT or biotechnology sectors or for general technology transfer) rather than identifying demand-side areas (e.g. those related to societal or developmental goals) or using incubators to support demand-driven startups. But the effective incubators that we assessed purposefully engaged in meeting societal and developmental demands that remained unmet. Unlike most incubators in India, many of the incubators we studied had purposefully-defined goals related to addressing market failures and meeting unmet demand (for societal goals that now map to SDGs).

CIIE and TBI-KEC emphasized heavily on addressing market needs. CIIE's market-oriented approach focused on assessing the viability of new products in underdeveloped sectors and in markets relevant for implementing societal goals. Through its various accelerator programs, CIIE used its understanding of business, markets, and market failures in sectors with high societal impacts (including agriculture, water, and clean energy). These accelerator programs aimed to find a product-market fit for advanced-stage innovators who had already developed prototypes or products by connecting them with potential stakeholders, customers, or investors. Similarly, TBI-KEC identified specific market demands and built its capabilities in electronics and information technologies to address these demands, ensuring success despite the challenges related to its limited resources outside of an urban cluster.

The life sciences incubators (C-CAMP and IKP) focused on developing STI in a particular industry, i.e., biotechnology. These incubators recognized the sector specific needs of biotechnology that are unmet by market forces—such as the need for resources in the form of special equipment and laboratory facilities, or the need for more time (compared to IT) to demonstrate market potential. The added resources needed for biotech startups are difficult to obtain worldwide but more so in developing countries, given that private investors prefer less capital-intensive, low-risk IT that can provide short-term returns. These incubators therefore addressed specific challenges for biotechnology and life science startups by offering targeted mentoring, equipment, technical expertise, and industry linkages.

### 4.3.3 Providing access to financing for startups

Incubators that were effective in implementing societal goals that now map to the SDGs actively facilitated early-stage investment in risky STI-based startups by directly administering funding for startups from government bodies (similar to most other public incubators in India, see section 4.2). But most notably, some incubators developed their own seed funds (besides helping attract external investment, as described in 4.2).

IKP, CIIE, and SV were eligible to directly invest in early-stage incubatee startups and were registered investors with Securities and Exchange Board of India (SEBI). CIIE ran its own seed fund, Infuse Ventures, to provide early stage funding for clean energy startups. IKP helped set up the India Innovation Fund for investing in early-stage startups in the life sciences. These investments mutually benefitted both incubators and startups—incubators with financial investments in startups were more deeply engaged in startup success while the reputation of these handful of effective incubators potentially also had a positive signaling effect in attracting later-stage private investments for the startups they were associated with. For example, in CIIE, eighty percent of the incubatees received follow-on financing from venture capital or angel investors within two years of incubation.

Other incubators facilitated startup financing by implementing public funding schemes related to DST, DBT, and others (Figure 3), besides engaging in the traditional incubator function of enabling external financial networks. TBI-KEC offered loans to startups through the DST and did not take any equity. Incubators in the life sciences (IKP and C-CAMP) administered BIRAC grants to incubatees along with administering and distributing other prestigious grants (e.g., grants from the Gates Foundation). SINE's location in Mumbai, that is both a financial hub and an emerging startup cluster, provided easy access to venture capital for incubatee startups, with more than 50 percent of incubatees receiving investments from angels, venture capital, and financial institutions.

### 4.3.4 Strengthening startup networks

Providing startup incubatees access to multi-faceted networks is a core incubator activity worldwide (2.2) but was particularly critical in India where market failures (2.1) made it harder for STI-based startups to have adequate resources or infrastructures, or links with potential investors and potential markets. These networks for knowledge (including technical, strategic, operational, and market knowledge), mentorship, finance, and private sector markets played a vital role in incubator effectiveness. TBI-KEC exemplified the importance of networks in a non-metropolitan region—incubatees benefited from the incubator's close ties with the local industry association (Coimbatore

District Small Industries Association) whose chairperson permanently served on the board of the incubator[15]. CIIE, IKP, and SINE utilized their networks to ensure meaningful mentorship for incubatees. C-CAMP emphasized on market linkages and exposure to business ideas for its startup incubatee scientists through mentor forums and events. Incubators like IKP and CIIE also offered startups access to knowledge networks by supporting business plans, technology licensing, compliance requirements, intellectual property, etc.

### 4.3.5 Managing incubators effectively

The implementation of public policy goals to expand the number of STI-based incubators (section 4.1) lacked direction on how these incubators would be managed, especially given the risks faced by STI-based startups and the market failures related to SDGs (Section 2.1). We found that a key determinant of effectiveness was the clear development of management direction that transpired from the ability of incubator leadership (managers, managing team, founders, or trustees of the incubators) to effectively develop critical incubator activities pertinent to STI-based startups and SDGs (described in 4.3.1-4.3.4) while supporting incubator operations.

The importance of managers and management strategy manifested in three ways. *One*, experienced and effective managers were better able to work cohesively with different government departments, innovators, academics, and local industries to deliver economic and sustainable development outcomes from incubator activities. The experience managers in the six more effective incubators had worked in the private or public sectors or had graduated from a top-ranked university with strong alumni networks. In contrast, managers in less effective incubators were often professors

[15] With the help of networks of the industry association, the first product launched out of the incubator was an industrial vacuum cleaner part for Hacko (a German company)

adept at scaling-up or managing technology but lacking in experience in high-risk activities or in connecting STI with market needs. *Two*, effective managers had the skills to manage incubator finances and develop business models that ensured long-term financial stability and a secure flow of income for the incubator. The five-year timeline of financial support from DST was at odds with supporting STI-based entrepreneurship, given that most early-stage startups need time to develop their products and to yield financial returns[16]. Consequently, while many incubators in the country struggled to be financially sustainable, the six incubators that we analyzed managed to generate revenues or to minimize costs using different business models. Incubators with their own seed fund (i.e., IKP, CIIE, and SV) charged a fund management fee to ensure sustainable revenue generation. IKP and C-CAMP complemented incubation activities with income generated through other sources—e.g., IKP charged a fee from foundations for managing grants and a fee from companies needing specialized biotech equipment; C-CAMP charged a licensing fee from users of its technology platforms. CIIE and SV also effectively mobilized private sector investments—e.g., CIIE's accelerator program (iAccelerator) had financial support from Microsoft; SV raised nearly INR 25 million of investment for the incubator from the private sector as matching funding. Incubators like TBI-KEC engaged in lean operations—i.e., less staff members with multi-faceted skills—to minimize costs and maximize revenues. *Three*, effective managers engaged in regular self-assessment exercises and in adjusting activities and outcomes to improve performance. While DST had no formal or standard reporting requirements on incubator performance, CIIE held regular internal reviews that served as guidelines for changes in its activities in accordance to demand-side market needs especially related to societal goals. In contrast, DBT rigorously monitored the performance of IKP and C-CAMP, resulting in efforts to develop metrics for self-assessment and to find opportunities for improvement in economic and sustainable development outcomes related to STI-based startups.

[16] For reference, most private equity firms that invest in risky ideas have a ten-year fund, and consequently, a ten-year investment horizon.

## 5. Strengthening STI-based incubators for implementing SDGs

Developing countries have recently renewed the emphasis on enabling STI-based startups with the help of various configurations of incubators. For example, in the early 2010s, led by their respective governments, "Start-up Brasil" and "Start-up Chile" launched incubator-like programs aiming to attract local and international entrepreneurs. China continues to fund its large incubator program, and the push to enable STI-based entrepreneurship, especially from universities, continues to come from the highest levels of the government (Lu, 2015). Furthermore, even in Kenya, Nigeria, Zimbabwe, and Rwanda where the emphasis on STI-based entrepreneurship is relatively new, planned incubator-like entities (hubs) are anticipated to play a tremendous role in the economic transformation of these countries despite various challenges associated with inadequate education and financing (Friederici, 2016; The Economist, 2017).

India has had a similar renewal of incubator activity in recent years that links to the SDGs, for example in its goals of increasing the number of existing incubators through national programs such as Startup India (70 new incubators)(UN India, 2019b, p. 9), Atal Innovation Mission (300 new incubators)(UN India, 2019a, p. 8) (see section 4.1).

The deepening focus on, and investments in, incubators in developing countries underlines the importance of learning from past incubator strategies, operations, and management. This is particularly important when the context of (and resources available to) incubators in developing countries can vary substantially, while meeting goals that depend on STI will require purposeful design (Barbero et al., 2014). Lessons learnt from an analysis of India—given its rich historical experience with STI-based incubators along with the need to manage economic development with other developmental challenges—are valuable not only for the country's own future efforts in extending incubation programs but potentially also for other developing countries facing comparable challenges or developing new programs.

Our analysis of India is one of the few comprehensive assessments of a complex incubator system that has existed for more than three decades(e.g., Lalkaka, 2002; Tang et al., 2013). Our findings thus create a foundation for new research centered around India and similar developing countries while filling a gap in the systematic understanding of past experiences as STI policies actively organize around SDGs. The Indian experience has shown both the potential and the limitations of incubators. In many cases, we have seen publicly-supported incubators as being remarkably effective in promoting STI-based entrepreneurship aimed at addressing developmental/societal challenges such as health and clean energy. But we also have seen that overall the public incubator programs, while successful in many ways, can also do better at supporting entrepreneurship more systematically and comprehensively.

Our analysis shows that incubators are effective when their activities go well beyond what has been commonly defined in traditional incubator literature centered around developed countries (described in 2.1). In India our findings suggest that incubators' extended activities include building human capacity, generating their own financing to support startups while also being the mandated channel for startups to get public financing (especially in the context of underinvestment by the private sector in STI for societal goods), and also helping incubatees better understand and connect to market and societal demands in underdeveloped markets. These extended activities are necessary because the context in which incubators operate in developing countries that are similar to India is vastly different from the countries that have been the subject of most incubator literature to date.

We outline below what we believe are some key steps that could help strengthen STI-based incubators, especially from the point of view of supporting SDG implementation efforts.

### 5.1 Building human capacity across the innovation system

#### Ensuring a pipeline of STI-based startups for incubators

Policymakers working on STI roadmaps for SDGs must ensure that capacity building for STI accompanies any efforts to develop additional STI-based incubators. Public policy must focus on systematically broadening and deepening the pipeline of STI-based entrepreneurs rather than relying on scattered measures in place in some well-performing incubators to do so (examples in 4.3.1). Building STI capacity at multiple levels in universities (i.e., in students, researchers, and faculty) is particularly needed in developing countries like India where universities are not the center of entrepreneurial activity (unlike in industrialized countries such as the United States) and universities are not strongly linked to innovation hubs (such as Silicon Valley).

We specifically suggest the following steps. *First*, strengthening science and engineering education is a necessary foundation for building human capacity to create STI entrepreneurs. We do understand that the tail cannot wag the dog, i.e., concerns about more effective STI-based entrepreneurship cannot drive higher education policy by itself. But it also is imperative for policymakers to realize that efforts to boost STI-based entrepreneurship eventually are dependent on the quality of graduates. *Second*, for researchers and faculty, exposure to the 'problem environment' (e.g. the technology and market needs associated with specific societal goals such as energy access) can develop avenues for linking STI with a larger set of SDGs. When combined with incentives to promote entrepreneurship (e.g., sabbatical year for entrepreneurial activities and flexible human resource policies), such activities can broad-base STI-based entrepreneurship. *Third*, at the university level, positive emphasis on startups can fundamentally change negative societal perceptions related to entrepreneurship (e.g., entrepreneurs-in-residence to act as role models to students[17], extending university's evaluation criteria to include university-based startups). In sum, effective capacity building in STI will be necessary for existing and new STI-based incubators to deliver outcomes related to the implementation of SDGs.

**Strengthening incubator and program managers**

Policymakers considering expansion of STI-based incubators in developing countries must also ensure that there is adequate managerial capacity to develop and lead incubator programs (4.2) as well as incubators (4.3.4). Both program managers and incubator managers play a key role in the implementation of STI policies through incubators[18]. Program managers can enforce clear hiring criteria for incubator managers—e.g., a combination of science and technology, business, and managerial capabilities—before funding new incubators while offering training and advisory support for the management team. Both incubator managers and program managers can help in optimizing operations (e.g., by conducting periodic evaluations, purposefully aligning operational goals with specific SDGs) or in optimizing incubator business models (e.g., by developing public-private or competitive tendering processes to leverage long-term financing in incubators[19], developing flexible sector-specific or performance-based financing programs for incubators). Managers therefore need technology as well as business experience for effective incubators and for the success of an incubator program.

[17] US universities engage in different activities to promote entrepreneurship among students by increasing interactions with successful entrepreneurs (see for example, MIT, 2016; Stanford, n.d.). For example, MIT invites successful alumni entrepreneur for one year (entrepreneur-in-residence) to guide students interested in founding startups in the developing world. Another example is the Mayfield Fellows Program at Stanford University that brings undergraduate students to Silicon Valley by offering them courses, mentoring and networking activities, and a paid internship at a startup in Silicon Valley.

[18] The importance of incubator managers for effective incubators has also been observed in China (Tang et al., 2014)

### 5.2 Organizing incubators around clear SDG targets

Policymakers can proactively use STI-based incubators as a tool to address market demand related to specific SDGs (see section 4.3.2). For example, governments could focus on designing incubators that have a purposeful objective of linking STI with market needs represented in SDGs (see examples in Table 1 related to energy access, health, sanitation, rural areas, water, agriculture, etc.). Such objectives could be enabled either through collaborations between government bodies or NGOs working with these issues and STI-based entrepreneurs[20], or through the procurement processes in government agencies (such as 'advanced market commitments') for technologies that have significant social benefit. Furthermore, given the systematic underinvestment by the private sector in STI for societal goods, governments could use incubators to target public funding towards early-stage STI-based startups related to SDGs (examples in 4.3.3). A recent example comes from

[19] Competitive tendering processes have been used to finance public-private incubators. For example, in Israel, the government implemented a public-private model for incubators by providing licenses to private equity, venture capital, angel investors, other industry, etc. through a competitive process. These incubator license holders financed 15% of the budget for a startup, and the government provided grants for the remaining 85%.

[20] For example, the Chicago CleanWeb Challenge hackathon provided city data to innovators and invited them to create technological solutions for environmental issues. In another example, the city government of Helsinki, helps startups by using technologies from cleantech startups including energy efficiency, low emissions public transport, waste management, district heating, water and air quality. Similarly, the local government in Sao Paulo, Brazil eased pre-qualification conditions for procurement tenders in favor of SMEs and startups. Sao Paulo also prioritizes procurement from startups as long as their bids are no higher than 10% of bids from non-startups.

India where a government agency partnered with a domestic philanthropic entity to establish a major clean energy incubator that aims to attract international entrepreneurs.

**5.3 Improving coordination within, and assessment of, 'incubation systems'**

Policymakers need to emphasize on systems-level coordination of existing incubator activities as they consider adding new incubators. Experiences from India show that a wide range of government programs and agencies engage in developing incubators, each with different policy goals—yet STI policy in the country has, so far, failed to learn systematically from past experiences, gather data, or organize around a coherent set of policy goals. For example, even the extensive efforts to analyze innovation in small businesses in the country through a survey of over 9,000 firms did not consider the effects of direct public support to incubators or incubatees on innovation outcomes (NSTMIS - DST GoI, 2014).

Coordination can help improve the effectiveness of individual government-led programs by minimizing overlaps and maximizing synergies, especially since our analysis shows that effective incubators tap into multiple public financial resources administered by different agencies (see Figure 3). This coordination is necessary not only between government agencies (and program managers) but also between incubator managers—e.g., through forums, sector-specific meet-ups, networks of incubators—and can help in systematic sharing of knowledge, experiences, and generation of new ideas and networks[21] among incubators and incubatees[22] (Cooper et al., 2012).

---

[21] For example, the Clean Energy Incubators Network in the US aims to highlight best practices on incubation techniques and clean energy technologies through workshops that bring together start-ups, incubators, investors, and industry participants working on clean energy.

[22] The Indian STEP and Business Incubator Association already organizes such meetings, but these meetings need expansion and could be formalized to require all managers.

Coordination can be improved in multiple ways: (a) A central body that takes on the role of coordinating various publicly-funded incubator programs (and therefore the public funding for incubatees) within a country could be very helpful for effective allocation of resources and the organization of STI-based incubation around common goals (for example in China[23]). (b) Top-down assessments of existing incubator programs run by such a body could also help define outcome metrics needed to assess whether sustainable development objectives have been met, followed by a systematic understanding of how to refine the overall approach towards incubation. (c) Given the importance of systematic data collection for program evaluation, this body could also mandate that all agencies involved in running incubators or financing their incubatees collect and submit data regularly, which can become the basis of a national database. (d) Similarly, assessments of (technological/sectoral or regional) innovation system dynamics in a country along with assessments of market needs related to SDGs could serve as valuable inputs in defining incubator strategy. These could be accompanied by sector- or region-specific support services that are imperative for STI-based startups—including professional technical assistance (e.g., through 'innovation vouchers' that cover costs of such assistance), legal support for intellectual property and patenting, market research, or access to centralized government laboratories that help in testing new technologies.[24]

---

[23] China's Ministry of Science and Technology tracked the progress of the China Torch Program (for science parks and incubators) allowing for periodic analysis and evaluation of the effectiveness of the program. The rich data allowed new research, for example, Hong and Lu, (2016) empirically found that professional technical services were particularly valuable to the incubatees.

[24] The validation of the technical performance of a new product by a government laboratory could help mitigate the perceived risk of investing in such a technology. For example, the Comprehensive Initiative on Technology Evaluation (CITE) is a USAID-funded program, where researchers at MIT develop consumer reports for new products (e.g., solar lanterns) provided by international aid agencies or private companies, to help consumers make informed choices of their purchases.

The absence of such systematic collection of data and analysis has been a major shortcoming of public policy in this area and urgently needs to be rectified. Since the activities related to data collection, coordination and assessment, while beneficial to all incubators and incubator agencies, are unlikely to be taken up by any individual entity, a top-down approach may be the only way to ensure action on this front. Policymakers with a system-level perspective are best positioned to set up a framework that manages such activities as part of STI policy implementation related to incubators. When setting up a performance monitoring mechanism, organizations such as the DST in India can learn from data gathering and monitoring experiences of other incubator programs, such as the Torch Program in China.

## 6. Conclusions

Publicly-funded incubators have been, and continue to be, a pivotal element in developing countries for promoting STI-based entrepreneurship. Our analysis of STI-based incubators in India shows that the goals for publicly funded incubators, even before the introduction of SDGs in 2015, often mapped to the sustainable development goals—most notably Goals 8 and 9. Now, as developing countries in particular organize public policy around the SDGs and increasingly value the role of STI to meet policy goals, lessons learned from past experiences of incubators can be valuable for effective design of incubators for implementing STI policy towards SDGs.

India's incubator experience suggests that incubators have been effective in their SDG related goals when their activities extended beyond 'traditional' incubator functions of providing infrastructure, networks, and services for startups that are commonly defined in the literature. These non-traditional activities include engaging in human capacity building activities for developing and identifying entrepreneurial talent, channeling public financing for startups but also supplementing that with their own seed funds, and in actively supporting areas related to specific SDGs (e.g., energy, health, etc.) that are beneficial to society and have a clear and high market demand but may not always be appropriately monetized by the private sector. At the root of these extended activities is the context under which incubators operate in developing countries that are similar to India. This points to the need for a redefinition of incubators and a shift in incubator theory that encompasses the outsized role of incubators in developing countries.

Future policy design on using STI based incubators to implement SDGs needs to consider the following aspects. *First*, incubators and public agencies involved incubator ecosystem should explicitly align their existing goals and objectives to the SDGs to set clear targets. *Second*, the 'incubator system' should be coordinated at national level to prevent proliferations of intermediaries or of redundant programs. *Third*, countries with an existing incubation system that do not have a robust monitoring of performance should develop such system before setting up new incubators. *Fourth*, countries which are at the early stage of setting up an incubation system should create a robust performance monitoring system right from the start of policy implementation. And *fifth*, a robust incubator policy is not enough—human capacity building, focused on the unique demand of STI, is needed at multiple levels including at the startup and at the incubator management level.

Overall, public policy for supporting STI-based entrepreneurship for implementing SDGs needs to focus on strengthening individual incubators as well as the 'incubation system'. Additional research is needed to develop frameworks and approaches for systematically tracking data on public funding of incubators and incubatees, for identifying relevant metrics of success in supporting startups, and for appropriate monitoring and evaluation of incubators and programs and how to map to the different SDGs.

**Acknowledgements:** The authors gratefully acknowledge the Indian Department of Science and Technology for its generous financial support through the IIT Delhi DST Center for Policy Research. The authors also gratefully acknowledge Professor Venkatesh Narayanamurti at the Science, Technology, and Public Policy (STPP) Program at the Belfer Center for Science and International Affairs at Harvard Kennedy School (HKS) for the support to Kavita Surana through a post-doctoral fellowship.

## Appendix

Table A1: Organizations involved with publicly-funded incubators in India that were interviewed for this study

| | **Organizations where interviews were conducted** |
|---|---|
| 1. | National Science and Technology Entrepreneurship Development Board (NSTEDB) |
| 2. | Department of Biotechnology (DBT) |
| 3. | Biotechnology Industry Research Assistance Council (BIRAC) |
| 4. | Indian STEP and Business Incubator Association (ISBA) |
| 5. | Centre for Innovation Incubation and Entrepreneurship (CIIE) |
| 6. | Centre for Cellular and Molecular Platforms (C-CAMP) |
| 7. | IKP Knowledge Park (IKP) |
| 8. | Technology Business Incubator at Kongu Engineering College (TBI-KEC) |
| 9. | Society for Innovation and Entrepreneurship (SINE) |
| 10. | Startup Village (SV) |
| 11. | Incubator with a host research institute |
| 12. | Incubator with a host large public university |
| 13. | Incubator with a host large technical university |
| 14. | Venture capital firm in India |
| 15. | Academia (researchers in innovation and entrepreneurship) |

Table A2: Illustrative list of semi-structured interview questions for incubator managers and staff. We used the term 'success' rather than 'effective' in the interviews for easier communication with the interviewees.

- What are the intended objectives for the incubator?
- What have been the envisaged activities to meet objectives?
- How have incubator activities changed over time?
- What have been the actual activities been carried out under those programs and by the incubators? And what might be the set of activities, a "successful" incubator must perform?
- What are the processes/determinants which lead to success or failure?
- How have incubators been able to strengthen the overall ecosystem?